\documentstyle[aps,prb,twocolumn,psfig]{revtex}
\begin{document}
\twocolumn[\hsize\textwidth\columnwidth\hsize\csname@twocolumnfalse\endcsname 
\title{$\pi-$shifted magnetoconductance oscillations 
in mesoscopic superconducting-normal-heterostructures}
\author{V.N.Antonov$^{1}$, H.Takayanagi$^{1}$, F.K. Wilhelm$^{2}$, and 
A.D.Zaikin$^{2,3}$} 
\address{$^{1}$ NTT Basic Research Laboratories, 3-1 Morinosato-Wakamiya, 
Atsugi-Shi, Kanagawa 243-01, Japan\\
$^{2}$ Institut f\"{u}r Theoretische Festk\"{o}rperphysik,
Universit\"{a}t Karlsruhe, D-76128 Karlsruhe, Germany\\
$^{3}$ P.N. Lebedev Physics Institute, Leninskii prospect 53,
117924 Moscow, Russia}
\maketitle
\begin{abstract}
Interference of proximity induced superconducting correlations in 
mesoscopic metallic rings is sensitive to the magnetic flux $\Phi$
inside these rings. This is the reason for magnetoconductance
oscillations in such systems. We detected experimentally and
explained theoretically a novel effect: the phase of these
oscillations can switch between 0 and $\pi$ depending on the
resistance of intermetallic interfaces and temperature. The effect is
due to a nontrivial interplay between the proximity induced
enhancement of the local conductivity and the proximity induced
suppression of the density of states at low energies.
\end{abstract}
]

In the past few years, mesoscopic superconductor-normal metal 
(SN) structures have attracted lots of theoretical and experimental interest 
\cite
{Lotsofstuff,VZK,GWZ,Yip,NazSt,Volkov,Petr,Courtois,Hartog,Morpurgo,WSZ,Gueron,Belzig}%
. Due to substantial progress in fabrication technology, nanoscale
structures with excellent inter-metallic interfaces were fabricated,
allowing to study the ``coherent'' dissipative conductance within the normal
metal, which is increased by the proximity effect. Although it was believed
that the proximity effect was already understood for many years, many
surprising observations such as non-monotonic ``reentrant'' conductance and
long-range coherent phenomena were observed experimentally and explained
theoretically \cite{GWZ,Yip,NazSt,Volkov,Petr,Courtois,Hartog,Morpurgo,WSZ}. 

It was demonstrated \cite{GK,Belzig} that in the presence of proximity
induced correlations a gap in the quasiparticle spectrum of a 
diffusive normal metal develops at low energies. The typical value of
this gap is set by the Thouless energy $E_{Th}={\cal D}/d^2$, where 
${\cal D}$ is the diffusion constant and $d$ is the relevant 
geometric size of the N-metal.
If a multiply connected SN-structure is put into a magnetic field, the
proximity induced superconducting correlations may interfere
destructively. As a result the proximity effect gets weakened and
the correlation-induced gap in the N-metal may be lifted \cite{GWZ}.
This opens a possibility to control the system conductance $G$ by an
externally applied magnetic flux $\Phi$. 

The effect of the magnetic field may be twofold. It is well known that 
in the presence of a good metallic contact between S- and N-metals
the proximity induced superconducting correlations always {\it increase}
the conductance of a quasi-one-dimensional SN structure as compared 
to that of the N-metal. Thus in this case destruction of proximity induced
correlations in the magnetic field leads to a {\it decrease} of
$G$ with increasing magnetic flux for sufficiently 
small $\Phi$. On the other hand, if one attaches tunneling electrodes 
to the normal metal with proximity induced correlations (e.g.
like it was done in \cite{Gueron}) the corresponding tunneling 
conductance is {\it suppressed} due to the presence of the proximity
induced gap in the spectrum of the N-metal. In this case lifting the gap
by applying the external flux $\Phi$ will {\it increase} the conductance
with increasing magnetic flux. 

In the present paper, we report experiments in a multiply-connected
SN-heterostructure, probed through highly transparent metallic contacts 
as well as through low transparent tunneling barriers. 
This allows us to study the two above phenomena and
their interplay within the same measurement. We observe oscillations of $G$
as a function of the flux $\Phi $, which can be $\pi $-shifted depending on
the probes. We also develop a theoretical description of the observed
phenomena. %
\begin{figure}
\centerline{\psfig{figure=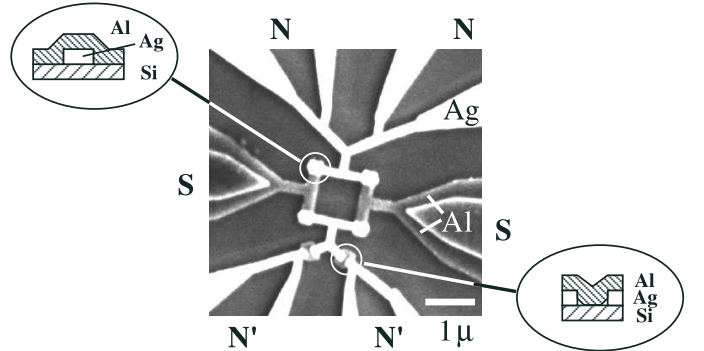,width=9cm}}
\caption{Scanning electron microscope
picture of the structure studied; $N-N$, $N^\prime -N^\prime$ and $S-S$
identify three different configurations (probes) for the resistance 
measurement. The schematic insets show the design of the NS interface
in the loop and the breaks in the $N^\prime -N^\prime$ probe.}
\label{sturcture}
\end{figure}

{\it Experiment}. The key point of our work is the design of the 
experimental sample. The
central object is a rectangular loop with opposite superconducting ($Al$)
and normal ($Ag$) edges (see Fig. 1). The superconductor induces
correlations into the whole sample due to the proximity effect.
The strength of the proximity effect can be controlled 
by the flux $\Phi $ through the loop, as it was discussed above.
 There are two probes $N-N$ and $N^{\prime }-N^{\prime }$ connected to
the normal edges of the loop. $N-N$ is brought into a good metallic contact to
the structure, whereas $N^{\prime }-N^{\prime }$ contains two breaks of
length $0.05-0.1\mu m$ near the measurement leads, which are bridged by the
aluminum strips. These breaks serve as artificial tunneling barriers
which decouple the silver wire from the external reservoirs. The 
extra resistance of 
$N^{\prime }-N^{\prime }$ varies between 1 and 5 $\Omega $ depending on the 
sample.
These resistive barriers play the key role in our experiment.
For its best performance it is desirable to have these 
barriers as high as possible. Unfortunately, their resistance 
cannot be done very large due to technological reasons. In the 
fabrication process, the aluminum strips are
defined simultaneously with the aluminum part of the loop. 
Since a strong proximity effect is needed, the contact resistance 
between the silver and the aluminum films should be low. To resolve 
the trade off
between these two requirements we define different overlapping areas for the
two metals in the loop and in the breaks ($0.2\times 0.25\mu m$ versus $%
0.1\times 0.1\mu m$ respectively). This helps to keep the $SN$ interface
resistance in the loop below $1\Omega $. Details of the fabrication process
can be found elsewhere\cite{Petr}. 
\begin{figure}
\centerline{\psfig{figure=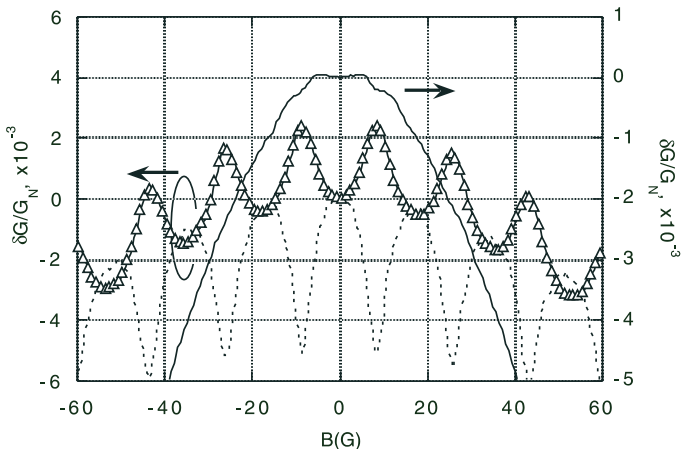,width=9cm}}
\caption{Magnetoconductance curves of the two probes: $N-N$ (dashed),
$N^\prime -N^\prime$ (triangles and solid). The curves are  taken at $T=0.3K$ (dashed
 and triangles) and at  $T=0.9K$ (solid). All curves are brought to zero at 
zero flux.}
\label{curves}
\end{figure}
The basic parameters for the silver film
are as follows: the thickness is $400\AA $, the width of the wires $0.1\mu m$%
, the sheet resistance $0.2\Omega $ and the phase breaking length $1\mu m$
at 0.3K. The sample has extra leads to measure the current and the voltage 
also inside the loop. We identify this as $S-S$ probe in Fig. 1. We study the
magnetoconductance of the normal, $N-N$ and $N^{\prime }-N^{\prime }$, and
superconducting, $S-S$, probes between $0.3K$ and $1K$. A lock-in technique
with a low frequency $f=183Hz$ and a measuring current of $1\mu A$ amplitude
is used to measure the conductance. The magnetic field was produced by a
superconducting solenoid and applied perpendicular to the sample plane. 

The magnetoconductance curves of the two probes are presented in
Fig. 2. In a small magnetic field, all curves show oscillations with $\Phi$
with a period equal to the flux quantum $\Phi _0=\hbar c/2e$ (the effective
loop area was defined by the center lines of the conductors). 
At low temperature the
oscillations in the low resistive $N-N$ probe have a maximum at $\Phi =0$ as
predicted theoretically. An overall envelope of the
curves is due to the pair-breaking effect of
the field penetrating into the normal conductors.
In the $N^{\prime }-N^{\prime }$-probe the phase of the oscillations {\it is
shifted} by $\pi $ as compared to that in the $N-N$ probe. This is
also in accordance with our theoretical predictions. At higher 
temperatures, a half-period component shows up
flipping the sign of the oscillations to that of the $N-N$ probe. This effect
is also not surprising: at higher $T$ the contribution of 
quasiparticles with energies above the gap 
$E \gtrsim E_{Th}$ becomes important (we
estimate $E_{Th}\sim 0.1K$ in the probes). For such $E$ the density of
states in the N-metal is not suppressed (on the contrary, it is even enhanced
in this energy interval) and the effect of the gap becomes unimportant.
Under these conditions the coherent contribution to the conductance 
dominates and determines the phase of oscillations. We will come
back to this point within our theoretical analysis presented below. 
\begin{figure}
\centerline{\psfig{figure=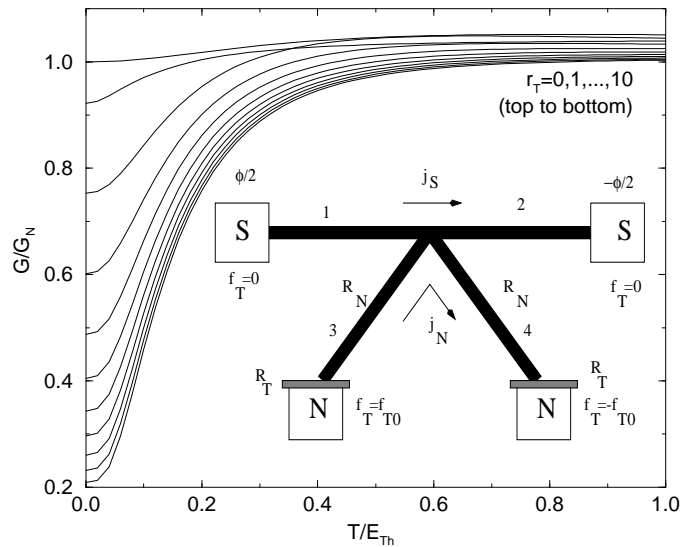,width=9cm}}
\caption{Linear conductance as a function of temperature at different 
tunneling resistances within the coherent/tunneling crossover at $\phi=0$. 
At $\phi=\pi$ we have $G=G_N$. The inset shows the simplified structure 
used for calculations. All arms 1--4 are assumed to have the same 
geometric parameters.}
\label{rt}
\end{figure}
{\it Theory}. Let us consider a structure which consists of an
SNS-junction with extra arms and normal reservoirs
attached to these arms via the tunnel barriers $R_{T}$ (see the 
inset of Fig. \ref{rt}). This structure captures all essential features 
of our system. A finite voltage drop exists between the normal reservoirs,
and the phase difference $\phi$ between the
two superconducting terminals can be related to the flux through the loop by
a simple gauge transformation  
$\phi =\pi \Phi/\Phi _0$.
In order to proceed we will use the standard formalism of the 
quasiclassical Green-Keldysh functions in the diffusive limit 
(see e.g. Refs. \onlinecite{LO,VZK,GWZ,NazSt}). The retarded
Green functions describing the spectral properties of the system
can be conveniently parameterized as $G^R=\cosh \alpha $, 
$F^R=\sinh \alpha e^{i\chi }$. These functions obey the Usadel
equation. In the normal metal it takes the form 
\begin{eqnarray}
{\cal D}  \partial^2_x\alpha &=&-2i\epsilon\sinh\alpha -({\cal D}/2)
\left(\partial_x\chi \right)^2\sinh2\alpha \nonumber\\
\partial_xj_\epsilon&=&0\;\;\; \mbox{,} \;\;\;
j_\epsilon=(\partial_x\chi )\sinh^2\alpha .
\label{retard}
\end{eqnarray}
No supercurrent is flowing in the arms $3$ and $4$, therefore in these
arms we have $j_\epsilon =0$. The equations (\ref{retard}) are
supplemented by the boundary conditions 
\begin{equation}
\label{eq:bcgreen}\left. \alpha \right| _S=-i\pi /2;\;\;\;\left. dr_T\partial
_x\alpha \right| _N=\left. \sinh \alpha \right| _N 
\end{equation}
where $r_T=R_T/R_N$ and $R_N$ is the normal state
resistance of the arm $3$ or $4$. The branching conditions at the
nodal point read\cite{GWZ} 
$$
\sum_{i=1}^4\vec \partial _i\alpha =0;\;\sum_{i=1}^2\vec \partial _i\chi =0 
$$
where $\vec \partial $ is the derivative in the direction of the respective
branch away from the node.

The dissipative current in our system 
\begin{equation}
j_N=\sigma _N\int_0^\infty d\epsilon \;D_T(\epsilon )\nabla f_T(\epsilon ) 
\end{equation}
is determined by the local spectral conductivity 
$D_T=\cosh ^2\hbox{Re}\alpha $ and the asymmetric component of the 
distribution function $f_T(\epsilon )=f(\epsilon )+f(-\epsilon )$. 
Solving the kinetic equations%
\cite{LO,VZK} in the side-arms, where no supercurrent is flowing, we obtain $%
f_T(\epsilon )$ and the linear conductance\cite{VZK,GWZ} 
\begin{eqnarray}
  \label{eq:conduc}
\frac{G}{G_N}&=&\int_0^d\frac{d\epsilon\;g(\epsilon)}{2T\cosh^2(\epsilon/2T)};\;
g(\epsilon)=\frac{1+r_T}{M(\epsilon)+r_T/\nu_T(\epsilon)}\\
M(\epsilon)&=&\frac{1}{d}\int_{\hbox{\scriptsize Arm 3}}
 \frac{dx}{\cosh^2(\hbox{Re} \alpha)};
\;\nu_T(\epsilon)=\left.\hbox{Re}(\sinh\alpha)\right|_{R_T}\label{DOS}
\end{eqnarray}
These results describe the interplay between two effects:
conductance enhancement due to the proximity induced correlations
coming from the S-metals (this effect is described by $M(\epsilon )$)
and its suppression due to the proximity induced gap (the latter
enters via the density of states $\nu $ reduced below the gap). 

{\it Results and discussion}. Let us consider the two opposite limits of small 
and large $r_T$. For $r_T\ll 1$ the tunnel barrier is irrelevant, 
and the voltage drops across the metal wire. Since the local conductivity
is enhanced due to the proximity effect everywhere in the
wire, its total conductance increases. Also the gap effect is softened: 
in this case a direct contact to a normal reservoir prevents from
forming a real gap in the spectrum and only a space-dependent pseudogap in the
density of states of the N-metals develops \cite{GWZ}. This pseudogap 
vanishes in the vicinity of a normal reservoir.
\begin{figure}
\centerline{\psfig{figure=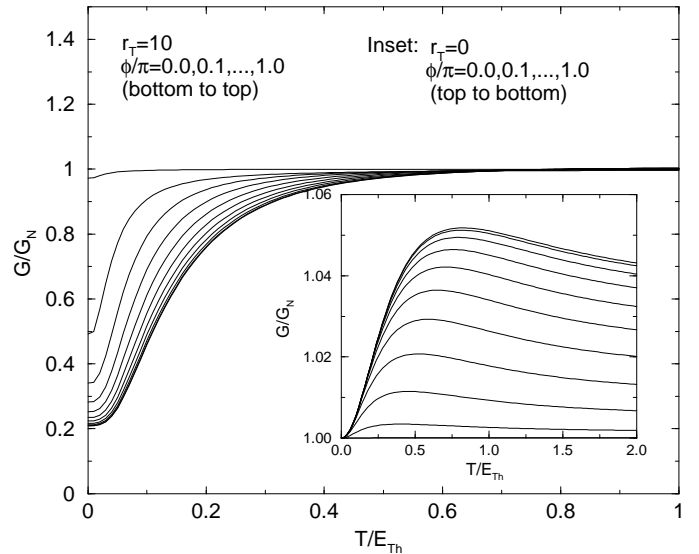,width=9cm}}
\caption{Linear conductance as a function of temperature in the 
tunneling-dominated regime; inset: coherent regime}
\end{figure}

At $r_T\gg 1$ the situation is entirely different. Now the main voltage
drop occurs at the tunnel barrier, and the enhancement of the local
conductivity of the N-metal becomes unimportant. In this case
the proximity induced gap in the density of states dominates the
system behavior leading to a {\it decrease} of the system conductance
at low energies. The coherent contribution to the conductance
described by $M(\epsilon)$ is relevant at higher $\epsilon$. 

The two regimes of small and large $r_T$ as well as the
crossover between them are described within our numerical analysis. 
At $r_T=0$, a typical non-monotonic ``reentrant'' behavior \cite
{GWZ,NazSt,Volkov} is reproduced, whereas at $r_T=10$ the 
tunneling dominates, see Fig. 3.%

It can be also seen that, in the latter case, the $G(\phi )$-relation is
highly non-sinusoidal and even has an edge around $\phi =\pi $. This
is also detected in our experiment, see Fig. 2.

The interplay between the  phase-coherent and the tunneling dominated 
contributions to the conductance for different values of $r_T$ and at different 
temperatures can be clearly observed in our results. In
Fig. \ref{rt} we show the total conductance at zero flux for different
values of the tunneling resistance. The conductance at $\phi=\pi$ is 
temperature independent and equals to $G_N$
in all cases. The intermediate curves illustrate that -- although at low
temperatures the conductance suppression due to the gap effect dominates
and therefore
$G<G_N$ -- at higher $T$ we have $G>G_N$ because for sufficiently large
$\epsilon$ the gap plays no role and the enhancement of
a local conductivity due to the proximity effect turns out to be more
important. 

For a quantitative comparison with the experiment, we can also generalize
our analysis allowing 
for a finite resistance $R_{SN}$ of the 
SN-interface in the 
loop. 
Comparing the corresponding numerical results with our experimental data for
$G(T)$ at $\phi=0$, we found that the best fit is achieved at
$R_{SN}\simeq 0.5\Omega $ in the
loop and $R_T\simeq 2.3\Omega $ at the barrier in the $N^{\prime}-N^{\prime }$ 
probe. These values are in reasonable agreement with the directly
measured values $R_{SN}=(0.7\pm 0.3)\Omega $ and $R_T\simeq 2.5\Omega $.
\begin{figure}
\centerline{\psfig{figure=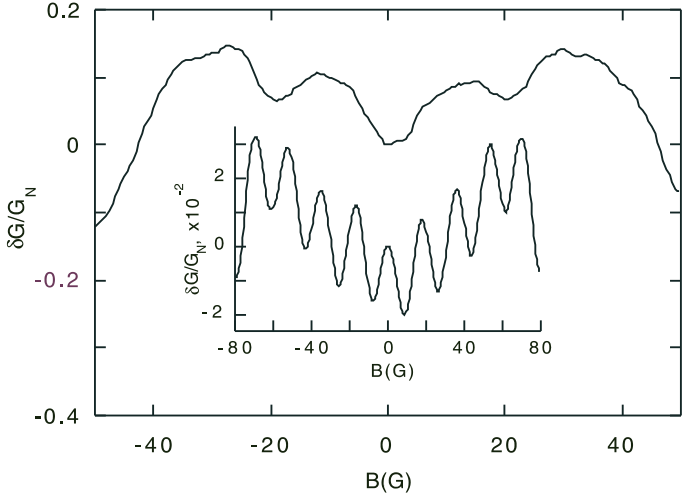,width=9cm}}
\caption{Magnetoconductance oscillations in the $S-S$ probe
 of the samples with $R_{SN}\simeq 4\Omega $ and $R_{SN}\simeq 0.7\Omega $ 
(inset). The curves are brought to zero at a zero flux.}
\label{SS}
\end{figure}
It was found in our experiment (see Fig. \ref{SS}) that the normal 
conductance of a metal between two superconductors
is also modulated by the flux and that conductance
oscillations have different phases, depending on the value of 
$R_{SN}$ similarly to the case of the $N-N$ channel considered above.  
At $R_{SN}\simeq 4\Omega $, we find a
minimum at zero-flux and at $R_{SN}\simeq 0.7\Omega $ a maximum. In the
former case the proximity effect in the side arms was extremely weak and $%
\delta G(\Phi )$ oscillations could hardly be detected. We can also add
that at low $T$ our treatment (developed for the $N-N$ channel) is 
not sufficient for a quantitative analysis of the charge transport in the
$S-S$-channel, in which case the Josephson current between
the two superconductors should also be taken into account. This current
may be strongly influenced by the nonequilibrium effects \cite{Morpurgo,WSZ}.
At higher $T\gg E_{Th}$ the supercurrent is exponentially suppressed 
whereas the coherent contribution to the conductance decays only
algebraically\cite{VZK,GWZ,Courtois}. In this regime the modulation
effects in the $S-S$ and $N-N$ channels are similar.

In summary, we detected a $\pi $-shift in the magnetoconductance oscillations
of a multiply connected mesoscopic normal metal-superconductor structure.
This shift is sensitive to the value of the resistance of 
intermetallic interfaces and temperature. The observed effect is
explained as a result of competition between the proximity induced
enhancement of the local conductivity and the proximity induced
suppression of the density of states at low energies.
A good agreement between our experimental and theoretical results
is found.

We wish to thank T. Claeson, T.M. Klapwijk, A.K. Geim, and D. Austing for
useful discussions and S.Gu\'eron for sending us her PhD-thesis. We are
grateful to Y.Tohkura and N.Uesugi of NTT Basic Research Laboratories for
their encouragement throughout this work. This work was supported by the DFG
through SFB 195, Graduiertenkolleg ``Kollektive Ph\"anomene im
Festk\"orper'' and the INTAS Grant 93-790-ext.

\end{document}